\documentclass[usenatbib,usegraphicx]{mn2e}
\newcommand{\ignore}[1]{}

\title[Photon counting strategies with L3CCDs]{Photon counting
strategies with low light level CCDs} 

\author[A. G. Basden et al.]{A. G. Basden,$^1$\thanks{E-mail:
abasden@mrao.cam.ac.uk} C. A. Haniff,$^1$ and C. D. Mackay$^2$\\
$^1$Astrophysics Group, Cavendish Laboratory, Madingley Road,
Cambridge CB3 0HE\\ 
$^2$Institute of Astronomy, Madingley Road, Cambridge CB3 0HA}

\begin{document}
\date{Released 2003 Xxxxx XX}

\pagerange{\pageref{firstpage}--\pageref{lastpage}} \pubyear{2003}
\label{firstpage}
\maketitle

\begin{abstract}

Low light level charge coupled devices (L3CCDs) have recently been
developed, incorporating on-chip gain.  They may be operated to give
an effective readout noise much less than one electron by implementing
an on-chip gain process allowing the detection of individual photons.
However, the gain mechanism is stochastic and so introduces
significant extra noise into the system.  In this paper we examine how
best to process the output signal from an L3CCD so as to minimize the
contribution of stochastic noise, while still maintaining photometric
accuracy.

We achieve this by optimising a transfer function which translates the
digitised output signal levels from the L3CCD into a value
approximating the photon input as closely as possible by applying
thresholding techniques.  We identify several thresholding strategies
and quantify their impact on photon counting accuracy and effective
signal-to-noise.

We find that it is possible to eliminate the noise introduced by the
gain process at the lowest light levels.  Reduced improvements are
achieved as the light level increases up to about twenty photons per
pixel and above this there is negligible improvement.  Operating
L3CCDs at very high speeds will keep the photon flux low, giving the
best improvements in signal-to-noise ratio.
\end{abstract}

This is a preprint of an Article accepted for publication in Monthly
Notices of the Royal Astronomical Society \copyright 2003 The Royal
Astronomical Society.

\section{Introduction}
Charge coupled devices (CCDs) are ideal detectors in many astronomical
applications. They are available in large-format arrays, have high
quantum efficiency (QE), a linear response, and, if cooled
sufficiently, a very low dark current. Their major shortcoming is
readout noise, i.e.\ the additional noise added by the on-chip output
amplifier, where the charge of the detected photo-electrons is
converted into an output voltage. Currently, the typical noise levels
achieved at slow readout rates (e.g. kHz pixel rates) are little
better than $\sim 2$ electrons per read \citep{jerram}. At the higher
readout rates (MHz pixel rates) often used, for example, for adaptive
optics and interferometric applications, far poorer performance is the
norm, with typical noise levels of $\sim 10-100$ electrons per readout
\citep{jerram}.

A novel solution to this problem, in which on-chip gain is used to
amplify the signal prior to readout, has recently been demonstrated by
E2V Technologies (formerly Marconi Applied Technologies,
\citet{jerram}). In this approach, an extended serial register is used
to allow electron avalanche multiplication so that a large mean gain
can be realised prior to a conventional readout amplifier.  Although
the effective gain can be very large, the detailed process by which
the signal is amplified is stochastic and so introduces additional
noise at the output. The effects of this noise, and its correction by
a judicious analysis of the output from a typical low light level CCD
(hereafter L3CCD) are the primary subjects of this paper.

The content of the paper is as follows. In Section 2 we describe how
the L3CCD works and develop a numerical and probabilistic model of the
gain mechanism.  In Section 3 we discuss techniques for analysing the
noisy output signal of an L3CCD and the implications for
photometric accuracy and signal-to-noise ratio (SNR), and we conclude in
Section 4.

\section{Characteristics of the L3CCD}
\subsection{Principles of L3CCD Operation}

\begin{figure}
\includegraphics[width=8.5cm]{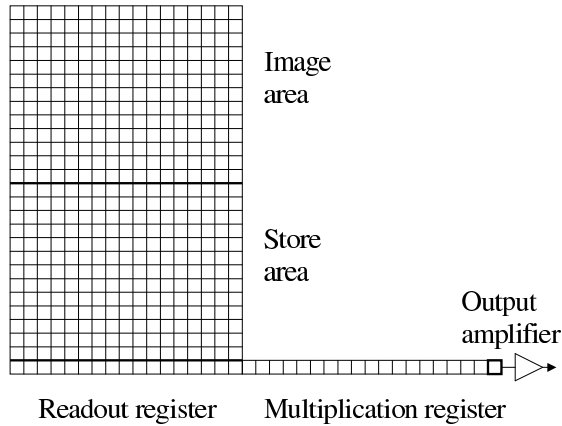} 
\caption{A schematic diagram of an L3CCD showing the multiplication register.}
\label{l3pic}
\end{figure}

The L3CCD architecture is similar to that of a normal CCD except that
it has extended serial register, called the multiplication register,
allowing for additional serial transfers before the signal reaches the
ouput amplifier (see Fig.~\ref{l3pic}). The electrode voltages in the
multiplication register can be adjusted so that avalanche
multiplication of the electrons occurs as they are moved through each
element of this part of the device.  At each step, the probability,
$p$, of producing an additional electron per initial input electron is
small --- typically this may be $0.01-0.02$ --- but the cumulative effect
of many transfers can be very large. For example, for a register
comprising $r = 591$ elements (the number of elements in the CCD65
from E2V Technologies), the mean gain, $g$, will be $\left(1+p
\right)^r = 6629$, when $p=0.015$.

\begin{figure}
\includegraphics[width=8.5cm]{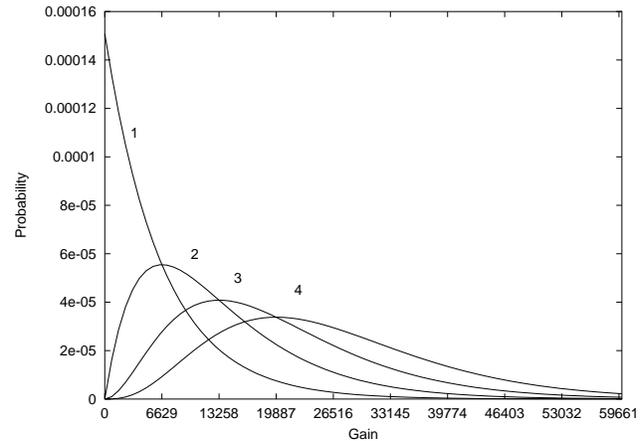} 
\caption{L3CCD output probability distribution for a given photon
input.  Inputs of $1-4$ photons are shown for a mean gain of 6629.
Note the significant overlap between different inputs which causes
uncertainties when estimating the input from the output.}
\label{probdistgraph}
\end{figure}

In general, the probability distribution of the output $x$, for an
input of $n$ (integer) photo-electrons, where the mean gain is $g$
will be given approximately by (Appendix A and Fig.~\ref{probdistgraph})
\begin{equation}
p \left( x \right) = \frac{x^{n-1}\exp(-x/g)}{
g^n\left(n-1\right)!}
\label{probdisteqn}
\end{equation}
when the photon input level is relatively small and the gain is large.
This distribution has a mean of $ng$ and a variance of $ng^2$ and at
high light levels is approximately Gaussian.  The signal noise
introduced by this multiplication process is independent of the photon
input.  The SNR of an L3CCD output is obtained by combining in
quadrature the noise due to the Poisson nature of light with the noise
from the multiplication process.  At a mean light level $\mu$ photons
per pixel, this gives an SNR equal to $\sqrt{\mu}/E$ where $E$ (the
excess noise factor, hereafter ENF) is equal to $\sqrt{2}$ for an
L3CCD output with large gain.  The nature of the multiplication
process means that in general there will not be a one-to-one mapping
between the number of electrons entering and leaving the
multiplication register (see Fig.~\ref{probdistgraph}), so that in
principle there will be always be some uncertainty when estimating the
input flux.  For a CCD with no gain, $E$ is equal to unity, and so to
achieve the same SNR when using a large gain, we will need to detect
twice as many photons, meaning that the multiplication process has
effectively halved the QE of the device.  This is a serious loss as
astronomers are happy to spend a lot of money on optical coatings to
increase system throughput by only a few percent, and do not want to
accept the loss in effective QE that the use of an L3CCD implies.  To
be able to increase the effective QE using signal processing is
therefore essential if L3CCDs are to be used effectively for
astronomy.  This is the main driving force behind our investigation,
which aims to use our additional knowledge of the system to allow us
to reduce the uncertainty in the measured fluxes.

\subsection{Reducing the noise}
\label{lllsect}
We consider processing strategies which allow up to minimize the
stochastic multiplication noise.  At very low light levels, much less
than one detected photon per imaging pixel, we are able to use the
L3CCD in a photon counting mode.  At these low light levels, we will
either get zero or one photons in a pixel per integration time.  This
will then result in either a signal much smaller than the amplifier
read noise (zero photons), or when a detected photo-electron is
amplified, a much larger signal.  Provided the mean multiplication
gain is much greater than the read noise, we can then treat any signal
above some threshold as having arisen from one photon event.
Replacing the range of signals we get from a pixel containing one
photon (curve 1 in Fig.~\ref{probdistgraph}) eliminates the variance
in output signal introduced by the multiplication process of the L3CCD
and increases the SNR to a value we expect from a conventional CCD
with negligible read-out noise.

At higher light levels this technique will not work, since coincidence
losses (more than one photon falling on a pixel, being interpreted as
one photon) will become increasingly important.  Nevertheless, the
idea of thresholding can still be helpful if we use more than one
threshold.  We can see how this might work as follows.  An L3CCD with
a mean multiplication gain of $g$ might give an output signal of $fg$.
This could be due to a single detected photon that has been amplified
by an unusually large amount.  Alternatively, if we set thresholds
$T_n$ and $T_{n+1}$ and estimate the flux to be $n$ detected
photo-electrons before multiplication if the detected signal satisfies
$T_n \leq fg < T_{n+1}$, then we will be able to maximize our chance
of estimating the flux correctly if we choose thresholds correctly.
It is the determination of these threshold levels and any corrections
that need to be applied in order to conserve flux which we now
consider, together with their impact on the L3CCD output SNR.

\subsection{Thresholding schemes}
We have considered many different signal-processing strategies, and
provide details of the most useful selection here.  We investigate
which of these provide the largest SNR improvement.  When
investigating these strategies, we assume that the data is first
thresholded at some level above the on-chip readout noise level
(typically $6\sigma$ where $\sigma$ is the RMS noise due to the
readout amplifier) so that amplifier read noise is negligible.
Following this, the strategies considered are:
\begin{description}
\item[1. Analogue:] The output signal is divided by the mean gain.
\item[2. Photon Counting (PC):] If the output signal is above a single
fixed noise threshold it is treated as representing one input photon.
\item[3. Poisson Probability (PP):] Threshold levels are set where the
probability of an output resulting from a mean input of $n$ Poisson
photons is equal to the probability of the output resulting from a
mean input of $n+1$ Poisson photons.
\end{description}

The Photon Counting (PC) thresholding strategy will underestimate flux
at light levels where there is a non-zero probability of more than one
photon being detected on a single pixel.  The Poisson Probability (PP)
strategy will overestimate the flux, particularly at low light levels,
and so a theoretically determined correction is necessary after the
detections, which we investigate in section 3.

\subsubsection{Threshold boundaries}
At a given mean light level, $\mu$ photons per pixel, the L3CCD output
with mean gain $g$ can be estimated by providing the photon input in
Eq. \ref{probdisteqn} with a Poisson probability distribution, giving:
\begin{equation}
p\left(x,\mu\right) = \sum_{n=1}^\infty
\frac{\exp\left(-\mu-x/g\right)\mu^n(x/g)^{n-1}}{g(n-1)!n!}
\label{sumprobdist}
\end{equation}
where $p\left(x,\mu\right)$ is the probability that the L3CCD output
will be $x$ when the mean light level is $\mu$, and the mean gain is
$g$.

\begin{figure}
\includegraphics[width=8.5cm]{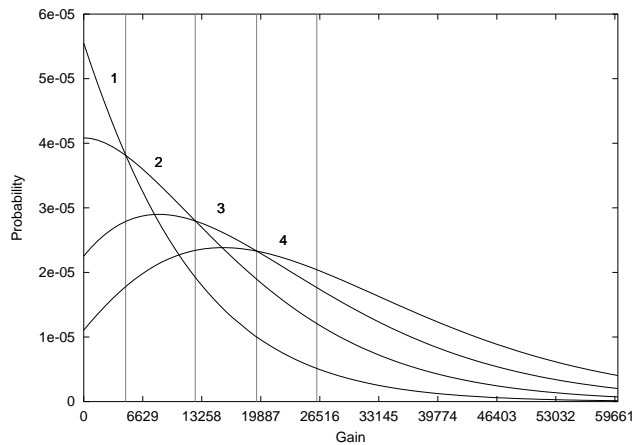} 
\caption{The L3CCD output probability distribution for mean (Poisson)
light levels of 1 (left curve) to 4 (right curve) photons per pixel,
as given by Eq.~\ref{sumprobdist}.  Threshold boundaries for the PP
thresholding strategy are placed where these curves cross.}
\label{poissonprobdistgraph}
\end{figure}

We use this distribution to determine threshold boundaries for the PP
thresholding strategy, which are given in Table~1.  These are placed
at the points where the probability of getting an output $x$ with a
mean light level $\mu$ is equal to the probability of getting the same
output $x$ with a mean light level of $\mu+1$ as shown in
Fig.~\ref{poissonprobdistgraph}, i.e. finding $x$ such that
\begin{equation}
p\left(x,a\right) =p\left(x,a+1\right)
\end{equation}
where the $a^\mathrm{th}$ threshold boundary is placed at position $x$, and
$p\left(x,a\right)$ is defined in Eq.~\ref{sumprobdist}.  This results
in threshold boundary positions independent of the mean light level.
The theoretically predicted mean flux and ENF arising from this
strategy is also calculated.

\begin{table}
\caption{Threshold boundaries in units of mean gain, for a PP
thresholding strategy.}
\begin{tabular}{clcl}\hline
Threshold & Boundary&Threshold&Boundary\\ \hline
1 &0.71 &7&6.97\\ 
2 &1.89 &8&7.98\\ 
3 &2.93 &9&8.98\\ 
4 &3.95 &10&9.98\\
5 &4.96 &11&11.0\\
6 &5.97 &$n\geq11$&$n$\\ \hline
\end{tabular}\\
\end{table}

\subsection{Threshold evaluation}
Monte-Carlo simulations were used to verify that our threshold
positions were chosen correctly, and that our calculations of photon
input estimation and ENF were correct.  Input photon streams were
generated for differing mean light levels assuming Poisson statistics,
and were injected into the first pixel of a multiplication register of
length $r$. The transfer of this signal to the next pixel of the
register was then computed assuming a probability $p$ that any given
photo-electron would be amplified to give 2 photo-electrons, and a
probability $1-p$ that the transfer took place with no
amplification. This process was then repeated a further $r-1$ times to
simulate the output expected for the initial input signal. In this
way, the expected output for L3CCDs with differing multiplication
probabilities and register lengths could easily be generated for
differing input light levels and threshold boundary positions.  Large
numbers of these output data sequences were then processed to verify
that the theoretical ENF and flux estimation calculations were
correct.

\subsection{Figures of merit}
In order to assess the performance of different signal-processing
strategies, the quality of the signal recovery was quantified using
both the ENF and the following misfit function, $M$:
\begin{equation}
 M =  \frac{\sum_\mathrm{i} \left( n_\mathrm{i} - y_\mathrm{i} \right)^2}
           { \sum_\mathrm{i} n_\mathrm{i} } , 
\label{misfiteqn}
\end{equation}
where $n_\mathrm{i}$ is the true input photon count and $y_\mathrm{i}$
the value of the input signal estimated from the thresholded L3CCD
output, $x_\mathrm{i}$. Data sequences comprising many tens of
thousands of signal values were generated so as to allow different
methods for generating the estimates $\{y_\mathrm{i}\}$ from the raw L3CCD
outputs $\{x_\mathrm{i}\}$.  The minimization of this misfit function
corresponds to the best input prediction.

An ideal detector will give a misfit of $M=0$, and a variance equal to
the variance of the input (Poisson) data, $\mu=\overline{n}$, with the
SNR equal to $\sqrt{\mu}$.  A non-ideal detector will have greater
dispersion in the output, resulting in a reduced SNR of
$\sqrt{\mu}/E$ where $E$ is the ENF.

\section{Discussion of improvements}
When investigating our L3CCD output thresholding strategies, it is
helpful to consider three different light level regimes.  The first of
these is when the mean light level is low, much less than one photon
per pixel per readout.  Secondly, we consider intermediate light
levels, with between about $0.5-20$ photons per pixel per readout, and
finally above this, high light levels are considered.  This separation
allows us to apply the different processing strategies able to
maximize the SNR at each light level.  Our results are independent of
multiplication register lengths typically found on L3CCDs (greater
than 100 elements), though a shorter register length will generally
give a slightly lower ENF (Eq.~\ref{matsuoeq}).

\begin{figure}
\includegraphics[width=8.5cm]{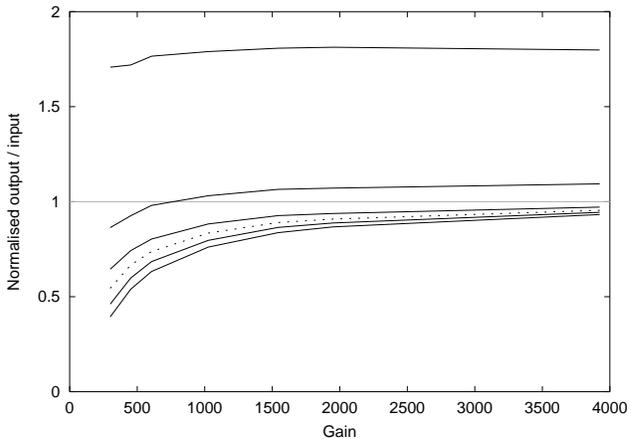} 
\caption{A graph showing the effect of on-chip readout noise for the
PP thresholding strategy as a function of mean gain, for different
noise threshold levels.  The upper curve has a noise threshold set at
$1\times\sigma$ above the noise level, $\sigma=50e^-$, $\sigma$
being the RMS noise, while lower curves set the noise threshold at
$2\sigma$, $3\sigma$, $4\sigma$ (dotted curve), $5\sigma$ and
$6\sigma$ above the mean noise.  A noiseless readout ($\sigma=0$)
would give a value of unity for all gains.}
\label{outputvgaingraph}
\end{figure}

\begin{figure}
\includegraphics[width=8.5cm]{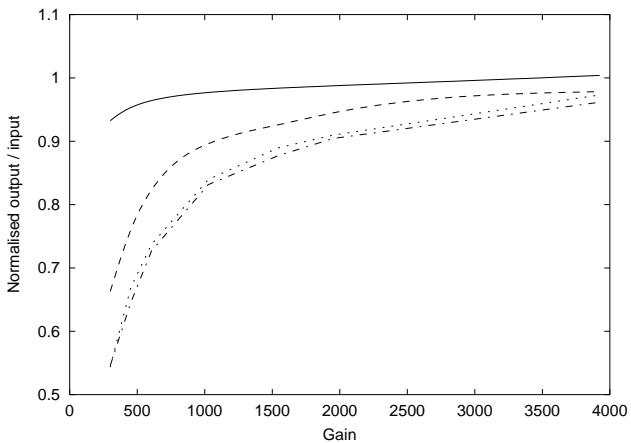} 
\caption{A graph showing the effect of on-chip readout noise for a PP
thresholding strategy when the noise threshold is set to $4\sigma$
above the mean noise level when $\sigma$, the RMS noise, is equal to
$50e^-$.  The upper (solid) curve is for 10 photons per pixel, lower
curves for 1 (dashed), 0.1 (dotted) and 0.01 (dash-dotted) photons per
pixel respectively. The effect of readout noise is minimized at higher
light levels and gains.  Curves are normalised to a value of unity
representing a noiseless readout.}
\label{outputfor4graph}
\end{figure}

As shown in Fig.~\ref{outputvgaingraph} and
Fig.~\ref{outputfor4graph}, the on-chip amplifier readout noise can
have a significant impact on the flux determination if the mean gain
is not much larger than the noise threshold level (below which any
signal is ignored), particularly at low light levels.  By setting the
noise threshold level at least $4\sigma$ above the mean on-chip
readout noise (assuming readout noise with an RMS of $\sigma$), and
using a mean gain at least ten times greater than the noise threshold
level, we are able to minimize the effect of the on-chip readout
noise.  As light level or gain are increased, the effect of readout
noise is reduced.

\subsection{Low light levels}
\begin{figure}
\includegraphics[width=8.5cm]{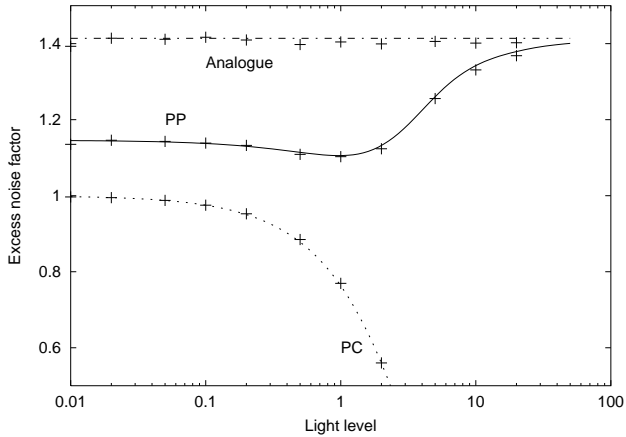} 
\caption{Excess noise factor as a function of light for the PP (solid
curve), PC (dotted) and analogue (dash-dotted) thresholding
strategies.  Curves are theoretical, while markers are from
Monte-Carlo simulation, and negligible readout noise is assumed. An
excess noise factor of $\sqrt{2}$ is equivalent to halving the QE of
the L3CCD.}
\label{enfvlightgraph}
\end{figure}

As discussed in section~\ref{lllsect}, if the mean light level is low
(much less than one photon per pixel per readout), we can use a PC
thresholding strategy, treating every signal above some noise
threshold as representing one photon.  This removes all dispersion
introduced by the multiplication process and effectively eliminates
any additional noise.  The signal-to-noise ratio from many such
samples scales as $\sqrt{n}$ as shown in Fig.~\ref{enfvlightgraph}.
The mean gain does not need to be accurately determined, but should be
kept well above the readout noise, as mentioned in the previous
section.  If this is not the case then some real signals will have
insufficient amplification and will be treated as noise, leading to
inaccuracies in flux estimation.

Coincidence losses will result in nonlinearities in the flux
prediction as the light level increases.  If the mean light level is
$0.2$ photons per pixel then two or more photo-electrons will be
detected on a pixel less than 2 per cent of the time, resulting in a
relatively small coincidence loss which can be corrected easily.
However, if the mean light level is $1$ photon per pixel, coincidence
losses are larger and we would estimate the light level to be only
$0.63$ photons per pixel.  This nonlinearity can be determined and so
we can correct the detected photon flux while this nonlinearity
remains small, for light levels up to about $1$ photon per pixel.  At
higher light levels, our estimated light level tends towards unity and
so we are unable to deduce the correct light level without much
uncertainty.

\begin{figure}
\includegraphics[width=8.5cm]{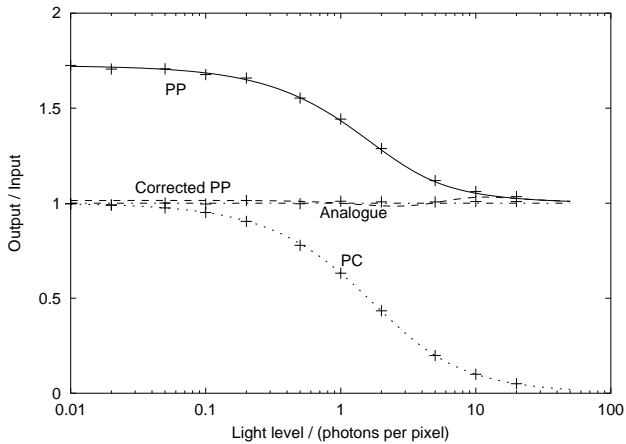} 
\caption{Ratio of estimated to true input for L3CCD output data using
the PP (solid curve), PC (dotted), corrected PP (using
Eq.~\ref{correction}, dashed) and analogue (dash-dotted) thresholding
strategies.  Curves are theoretical, while markers are from
Monte-Carlo simulation and negligible readout noise is assumed
($g\gg\sigma$).  We see that flux is overestimated at low light levels
using a PP strategy, but that this nonlinearity can be corrected for.}
\label{outputvlightgraph}
\end{figure}

Our PC thresholding strategy may be applied here, but will
overestimate the flux, since there is a significant probability that
the output from a single photon input will be placed into the second
(or greater) threshold bin.  Fig.~\ref{outputvlightgraph} shows the
size of the error in flux estimation for our PC and PP thresholding
strategies as a function of light level, and hence the nonlinearity
correction that should be applied after data from many frames has been
thresholded in this way.

\subsection{Multiple thresholding at intermediate light levels}
At intermediate light levels up to about twenty photons per pixel, we
cannot use a PC thresholding strategy as coincidence losses become
large.  However, as discussed previously it is still possible to
process the output, reducing the ENF.  Our PP processing strategy can
be applied at any light level though is most advantageous in this
light level regime, and gives decreasing improvements up to about $20$
photons per pixel per frame. Above this, the photometric accuracy and
ENF are indistinguishable from those obtained using the analogue
processing strategy.

Threshold boundaries from the PP processing strategy are independent
of light level.  Summing thresholded output signal values and applying
a nonlinearity correction for light level provides us with a good
estimate of the flux.  Fig.~\ref{outputvlightgraph} shows the
nonlinearity correction which should be applied.  Without this, the
flux will be overestimated at lower light levels since there is always
a significant probability that the output from a single photon will be
interpreted as two or more photons.

We find that theoretical results and those from our Monte-Carlo
calculations agree almost perfectly, as would be expected, as shown
for example by Fig.~\ref{outputvlightgraph}.

\subsection{High light levels}

At high light levels, the input photon (Poisson) distribution becomes
symmetrical about the mean light level, having the form of a Gaussian.
The multiplication noise distribution also tends to a Gaussian, and
so the L3CCD output distribution is Gaussian.

\begin{figure}
\includegraphics[width=8.5cm]{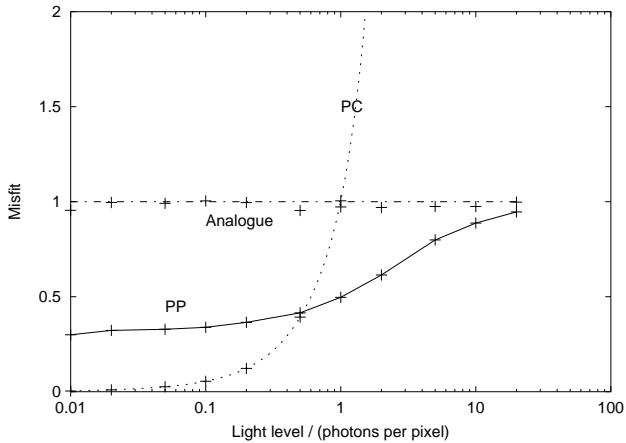} 
\caption{Misfit as a function of light for the PP (solid curve), PC
(dotted) and analogue (dash-dotted) thresholding strategies.
Negligible readout noise is assumed.  We see that a PC strategy should
not be used at light levels greater than about one photon per
pixel. pics2/misfitvlight.eps}
\label{misfitvlightgraph}
\end{figure}

We treat the output as we would a conventional CCD, simply dividing by
the mean gain, using the analogue processing strategy.  This does not
remove any of the dispersion introduced by the multiplication process,
giving an ENF, $E$, of \citep{matsuo}
\begin{eqnarray}
\label{matsuoeq}
E^2 & = & \frac{1}{g}+\frac{2}{g^{1/r}}-\frac{2}{g^{1+1/r}}\\ 
& \approx & 2\nonumber
\end{eqnarray}
where $r$ is the multiplication register length, $g$ the mean gain,
and the approximation is valid when $r$ and $g$ are large, as is usual
for an L3CCD.  The SNR is then $\sqrt{n/2}$, effectively halving the
QE, and the misfit function (Eq. (\ref{misfiteqn})) is also greater
than for other thresholding strategies, as shown in
Fig.~\ref{misfitvlightgraph}.

This analogue processing technique may be applied at any light level,
though due to the large ENF, other methods can give an improvement in
SNR, particularly at light levels below about 20 photons per pixel per
readout.  At very high light levels, where the mean light level is a
few times the square of the on-chip amplifier read noise of the L3CCD
(typically $50-100\mathrm e^-$ for fast readout, \citet{jerram}), the
multiplication gain can be turned off, and then no statistical noise
is added, giving Poisson shot noise scaling as $\sqrt{n}$.  This is
the mode in which a conventional CCD is used.

\subsection{Excess noise factors}
An ENF of $\sqrt{2}$ effectively halves the QE of the CCD, which is a
serious loss.  Fig.~\ref{enfvlightgraph} compares the ENF for various
thresholding modes, as a function of light level.  We can see that a
combination of PC and PP thresholding strategies allows us to reduce
the ENF for light levels up to about twenty photons per pixel.

It is interesting to note that the ENF tends towards zero for the PC
strategy at high light levels.  This is because we always interpret
every signal as one photon, and so the result has no noise.  However,
in this mode, we are unable to predict the input flux, and so it is
useless at high light levels, as seen from the misfit function
(Fig.~\ref{misfitvlightgraph}).

For the PP thresholding strategy, the ENF tends towards $\sqrt{2}$ as
the light level increases, and above about twenty photons per pixel we
see that there is little advantage in thresholding, with the raw L3CCD
output giving similar noise performance.  However, thresholding does
lead to a significant improvement in noise performance, reducing the
ENF up to twenty photons per pixel.

We also see that the misfit function remains low when we apply these
thresholding schemes (Fig.~\ref{misfitvlightgraph}) using a
combination of the PC and PP thresholding strategies.  The PC
thresholding strategy only performs well for light levels up to about
0.5 photons per pixel, and should not be used above 1 photon per
pixel.  

\subsection{Photometric correction}
Fig.~\ref{outputvlightgraph} shows that a nonlinearity correction for
flux will be required during post-processing for light levels greater
than about 0.1 photons per pixel when using a PC thresholding
strategy, and at light levels less than about 20 photons per pixel
when using a PP thresholding strategy.  We can correct the flux for a
PP thresholding strategy approximately according to
\begin{equation}
I_\mathrm{corrected}\approx 
\frac{I_\mathrm{est}}{1+0.7\exp(-I_\mathrm{est}/3)}
\label{correction}
\end{equation}
where $I_{\mathrm est}$ is the result of the initial thresholding and
summation process.  The result of this correction is shown in
Fig.~\ref{outputvlightgraph}.  Similarly, the flux for a PC
thresholding strategy can be corrected according to
\begin{equation}
I_\mathrm{corrected} = - \ln \left( 1-I_\mathrm{est} \right)
\label{PCcorrection}
\end{equation}
though the error in corrected flux becomes large as
$I_\mathrm{est}\rightarrow 1$.

\subsection{Lucky imaging}
As an example of where thresholding techniques can be used, we
consider the Lucky Imaging technique \citep{cdm}, as used in astronomy
to overcome atmospheric effects on medium sized telescopes.  Snapshot
images are taken with very short exposure times (of order 10-30ms).  Many
such images are taken, and the images with least atmospheric smearing
are kept and added together after centroiding.  An L3CCD is required
since the light levels will be very low. 

We can threshold each data frame using both the PC and PP thresholding
strategies and summing with previous frames immediately (in parallel
since these thresholding schemes are applied after detection). At the
end of the observing run, these two images can then be combined after
applying the nonlinearity correction for flux, depending on whether
the signal on each pixel is low or high.

\subsection{Sources of errors}
Apart from when using the photon counting (single thresholding)
strategy, we require knowledge of the mean gain.  This can be
controlled to about 1 per cent \citep{mackay} which, with gains of order
1000, requires millivolt stability for the clock-high L3CCD electrode
voltage.  Errors here will only have a small effect on our estimates
at low light levels and at higher light levels, the flux estimate
error will be proportional to the error in gain.

\section{Conclusions}
We have studied the stochastic gain process of L3CCDs and
characterised the noise added by this multiplication process.  For
different signal level regimes, we have investigated the best
processing strategy for the L3CCD output with regard to estimating the
true photon input.

In summary, we find that:
\begin{enumerate}
\item Single thresholding of the L3CCD output can be applied
accurately for photon rates up to about 0.5 photons per pixel per
readout with a small nonlinearity correction.
\item Multiple thresholding (binning) of the L3CCD output can be
applied at any light level, and is most advantageous with light levels
between 0.5-20 photons per pixel per readout.  This can reduce the
excess noise factor introduced by the multiplication process from
$\sqrt{2}$ to $1.1$ at light levels of about one photon per pixel.
\item Using a combination of single and multiple thresholding
strategies leads to further improvement, decreasing the excess noise
factor to unity at lower light levels.
\item If the gain is not known accurately, threshold boundaries will be
chosen wrongly.  However, the gain can be controlled to about 1 per cent,
so this effect is small.
\end{enumerate}

Our recommendation when using L3CCDs at low light levels (less than
0.5 photons per pixel per readout) is that a single threshold
processing strategy on the L3CCD output should be used.  At higher
light levels up to about 20 photons per pixel, threshold boundaries
placed with the PP thresholding strategy should be used.  A
nonlinearity correction should then be used, leading to correct flux
estimation and an improvement in SNR performance from $\sqrt{n/2}$ to
$\sqrt{n}/1.1$ in the best case, without requiring an initial estimate
of the mean light level.  At high light levels greater than about 20
photons per pixel, using the raw output does not lead to worse SNR
performance than that obtained with other thresholding techniques.

Since L3CCDs provide the best SNR performance at low light levels
using a single threshold, we recommend that if possible, they are
always used in this regime, increasing the frame rate if necessary to
keep the number of photons per pixel low ($<0.5$).  A new controller
being developed for L3CCDs will allow pixel rates of up to 30MHz,
allowing the potential of L3CCDs to be maximized.

\section*{Acknowledgements}
AB would like to thank Bob Tubbs for useful comments and discussions.

\appendix
\section{Probability Distribution}

\citet{matsuo} give the output probability distribution
for a electron multiplication device (e.g.~an L3CCD), with $r$
elements in the multiplication register, and a probability $P$ of
producing an extra electron at each stage, for a single photon input,
as

\( \begin{array}{lclrr}
p_{r}(x) & = & (1-P)p_{r-1}(x) & \\
& & + P\sum_{k=0}^x p_{r-1}(x-k)p_{r-1}(k), 
& x,r\geq 1 \\ 
p_r(0) & = & 0, &  r\geq 1 \\
p_0(x) & = & \delta_{1,x}, & r\geq 1 
\end{array} \)

If $r$ is large and $P$ is small, we find that this can be
approximated to an exponential distribution:
\begin{equation}
P_1 (x) = g^{-1}\exp(-xg^{-1})
\label{probdist1eqn}
\end{equation}
where $P_1(x)$ signifies the probability of an output $x$ for a single
photon input, with mean gain $g$.  This gives $<x>= \sigma_x^2 =
g$, as expected.

To generate the probability distribution for two input electrons, we
simply take the convolution of Eq.~\ref{probdist1eqn} with itself.
This gives:

\begin{eqnarray}
P_2(c) & = & \sum_{x=2}^{c}
       g^{-2}\exp(-xg^{-1})\exp(-(c-x)g^{-1})\nonumber \\
& =  & g^{-2}(x-1)\exp(-xg^{-1})
\end{eqnarray}

Likewise, the probability distributions for larger input electron
counts can be derived using

\begin{equation}
P_n(c) = P_{n-1}(x) \ast P_1(x) = \sum_{x=n}^c
P_{n-1}(x)\times P_1(c-x)
\nonumber
\end{equation}

where $\ast$ represents convolution.  This gives:
\begin{eqnarray}
P_3(x)&=&\frac{\left(x-2\right)\left(x+1\right)\exp(-xg^{-1})}{g^3 2!}
\nonumber \\
P_4(x)&=&\frac{\left(x^3-7x-6\right)\exp(-xg^{-1})}{g^4 3!}\\
P_5(x)&=&\frac{\left(x^4+2x^3-13x^2-278x+936\right)\exp(-xg^{-1})}{g^5 4!}\nonumber
\end{eqnarray}

where $P_n(x)$ is valid from $x\geq n$ (and is in fact zero at
$x=n-1$).  We can see that for moderately large $x$ (and $x$ usually will
be large over most of the distribution, since the gain is large) we
can simplify to give a general probability distribution:

\begin{equation}
P_n(x)=\frac{x^{n-1}\exp(-xg^{-1})}{g^n \left(n-1\right)!}
\end{equation}
which has an expectation $ng$ and variance $ng^2$, and fits the actual
distribution almost perfectly.  Variations on this are possible, for
example taking more care at small $x$, though since the overall
differences are small, the simplified version is used here.  This
approximation is not valid for large $n$, and a Gaussian distribution
with the same mean and variance should be used instead.

\bsp
\label{lastpage}
\end{document}